\documentclass[namedreferences]{solarphysics}
\usepackage[optionalrh,solaenum]{spr-sola-addons} 
\usepackage{graphicx}        
\usepackage{color}           
\usepackage{url}             
\usepackage{subfig}

\newcommand{\degr}{^{\circ}}

\newcommand{\feviii}{{Fe~{\sc viii}\,}}

\newcommand{\fex}{{Fe~{\sc x}\,}}
\newcommand{\fexii}{{Fe~{\sc xii}\,}}
\newcommand{\fexiii}{{Fe~{\sc xiii}\,}}

\newcommand{\fexv}{{Fe~{\sc xv}\,}}

\newcommand{\nixvi}{{Ni~{\sc xvi}\,}}
\newcommand{\heii}{{He~{\sc ii}\,}}

\newcommand{\arcsec}{$^{\prime\prime}$}

\newcommand{\kms}{km~s$^{-1}$}

\begin{document}

\begin{article}

\begin{opening}

\title{Study of the three-dimensional shape and dynamics of coronal loops observed by \textit{Hinode}/EIS.\\ {\it Solar Physics}}

\author{P.~\surname{Syntelis}$^{1}$\sep 
        C.~\surname{Gontikakis}$^{1}$\sep
        M.K.~\surname{Georgoulis}$^{1}$\sep        
        C.E.~\surname{Alissandrakis}$^{2}$\sep
        K.~\surname{Tsinganos}$^{3,4}$
       }
\runningauthor{Syntelis P. et al.}
\runningtitle{On the shape of active region coronal loops observed by \textit{Hinode}/EIS.}

   \institute{$^{1}$ Research Center for Astronomy and Applied Mathematics, Academy of Athens,\\ Soranou Efesiou 4, 11527, Athens, Greece\\
                     email: \url{psyntelis@phys.uoa.gr}\\ email: \url{cgontik@academyofathens.gr}\\
              $^{2}$ Section of Astro-Geophysics, Department of Physics, University of Ioannina, 45110 Ioannina, Greece
                     email: \url{calissan@cc.uoi.gr} \\
              $^{3}$ National Observatory of Athens,\\ Lofos Nymphon, Thission 11810, Athens Greece \\
	      $^{4}$  Section of Astrophysics, Astronomy and Mechanics Department of Physics, University of Athens, Panepistimiopolis 157 84, Zografos, Greece
                     email: \url{tsingan@phys.uoa.gr} \\
             }

\begin{abstract}
We study plasma flows along selected coronal loops in NOAA Active Region 10926, observed on 3 December 2006 with  \textit{Hinode's EUV Imaging Spectrograph} (EIS). 
From the shape of the loops traced on intensity images and the Doppler shifts measured along their length we compute their three-dimensional (3D) shape and plasma flow velocity using a simple geometrical model. This calculation was performed for loops visible in the \feviii\ 185~\AA, \fex 184~\AA, \fexii\ 195~\AA, \fexiii\ 202~\AA, and \fexv\ 284~\AA\ spectral lines. In most cases the flow is unidirectional from one footpoint to the other but there are also cases of draining motions from the top of the loops to their footpoints. Our results indicate that the same loop may show different flow patterns when observed in different spectral lines, suggesting a dynamically complex rather than a monolithic structure.
We have also carried out magnetic extrapolations in the linear force-free field approximation using SOHO/MDI magnetograms, aiming toward a first-order identification of extrapolated magnetic field lines corresponding to the reconstructed loops. In all cases, the best-fit extrapolated lines exhibit left-handed twist ($\alpha < 0$), in agreement with the dominant twist of the region.\end{abstract}
\end{opening}

\section{Introduction}
     \label{sec:Introduction}
Solar telescopes provide us with two-dimensional projections of coronal structures on the plane of the sky. The measure of the 3D geometry of coronal loops is important for  understanding these structures. Reconstructions of a loop's true shape have been attempted via a variety of methods, such as stereoscopy or models that make assumptions for the loop shape, flow etc \cite{Loughhead1983,Berton1985,Aschwanden1999,Nitta1999}. Moreover, the STEREO space mission, with its twin telescopes in two different positions in space, is dedicated to the solar coronal stereoscopy by combining simultaneous EUV images from two different lines of sight \cite{Aschwanden09,Aschwanden11}. The 3D geometry of active-region loops, computed with STEREO images, was used to constrain the magnetic field topology computed with non-linear force-free extrapolation methods \cite{DeRosa2009}. 
\begin{figure}
\centerline{
\includegraphics[width=0.7\textwidth]{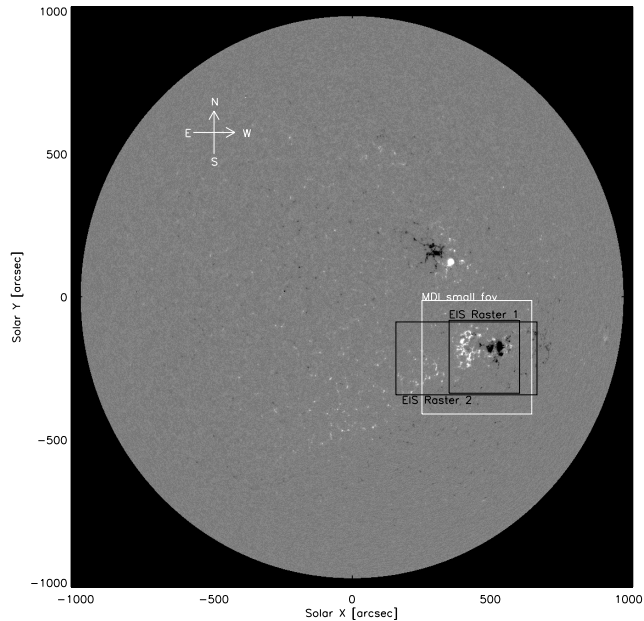}
}
\caption{Full disk image recorded by SOHO/MDI on 3 December 2006, at 20:51~UT. EIS rasters 1 and 2 field of views are represented with dark frames around AR 10926. The white frame shows 
the part of the MDI magnetogram used for the magnetic field extrapolation.}
  \label{fig:MDIfulldisk}
\end{figure}
To interpret the active-region Doppler maps computed with the data from instruments such as \textit{Hinode}/EIS and to understand the plasma flows along coronal loops, one also needs to know the loops' 3D geometry. \inlinecite{DelZanna2008}, using \textit{Hinode}/EIS observations of NOAA AR 10926, studied the behavior of the line of sight velocity, along loops and weak emission regions of the active region, at different plasma temperatures. 

In this work we analyze the same active region and we study the flows along selected loops for which we were able to reconstruct their 3D geometry. Our method, introduced by \opencite{Alissandrakis2008} (Paper~I) also uses the Doppler shifts measured along coronal loops in this reconstruction. Based on the analysis of Paper~I we extend the study to more loops and a wider range of formation temperatures in this work.

\section{Observations}
      \label{sec:Observations}
NOAA AR 10926 was observed on 3 December 2006, at (395\arcsec, -198\arcsec ) from disk center (see Figure~\ref{fig:MDIfulldisk}). During that day, \textit{Hinode's  EUV Imaging Spectrograph} (EIS) performed two rasters of the AR, recording spectral lines of \feviii\ 185~\AA, ($10^{5.8}$~K), \fex\ 184~\AA, ($10^6$~K), \fexii\ 195~\AA, ($10^{6.1}$~K), \fexiii\ 202~\AA, ($10^{6.2}$~K), and \fexv\ 284~\AA, ($10^{6.3}$~K). These formation temperatures have been published by \opencite{Young2007}. Raster~1 was obtained from 15:32:19~UT to 17:46:31~UT with a 256\arcsec\ $\times$ 256\arcsec\ field of view (FOV), while raster~2 was obtained from 19:15:12~UT to 23:44:09~UT and covered a 512\arcsec\ $\times$ 256\arcsec\ FOV Both rasters were scanned from West to East (see Figure~\ref{fig:MDIfulldisk}) with a 30~s exposure time for each slit position, and a 1\arcsec\ spatial pixel size. We also used a timeseries of TRACE 171~\AA\ filtergrams, recorded from 19:10:49~UT  to 23:49:06~UT, with 0.5\arcsec\ spatial pixel size and a time cadence of 1 minute. Magnetic field extrapolations were performed using a SOHO/MDI magnetogram. The MDI full disk line-of-sight magnetogram (Figure~\ref{fig:MDIfulldisk}) has a rough pixel size of 1.9\arcsec\ but covered the entire active-region field-of-view and was recorded on 3 Dec. 2006 at 20:51:01~UT, during raster~2. SOT provided a vector magnetogram with a pixel size of 0.15\arcsec\ in a FOV of 316\arcsec\ $\times$ 157\arcsec, 
pointing at (447\arcsec,-171\arcsec), that was taken on 4 Dec 2006, from 10:07:00~UT  to 13:03:36~UT.
\begin{figure}
\centerline{
	    \includegraphics[width=1.\textwidth]{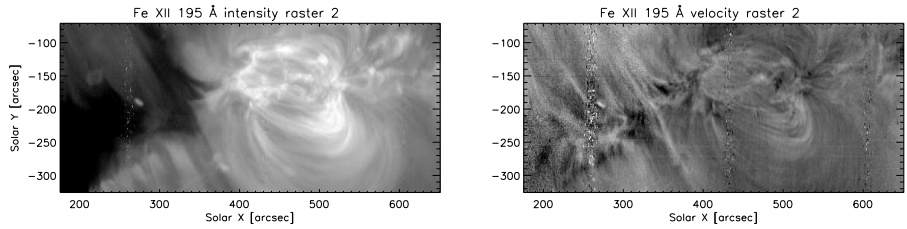}
	    }
\caption{Left: EIS intensity image of the active region in \fexii\ 195~\AA\ for raster~2. Right: The corresponding EIS dopplergram. For the dopplergram, white/black colors
corresponds to red/blue shifts. Doppler shifts are in the range of -20 to 20~\kms.}
  \label{fig:intenstiy-velocity-plots}
\end{figure}


\section{Data processing and loop reconstruction} 
      \label{sec:data_treatment}

The EIS data were treated following standard procedures. After correcting for electron spikes, pedestal and dark current, hot pixels, slit tilt and thermal drift, we performed Gaussian fits on all individual line spectral profiles to compute Dopplergrams and intensity images (see Figure~\ref{fig:intenstiy-velocity-plots}). \feviii\ 185~\AA\ is blended with \nixvi\ 185.23~\AA\ ($T=10^{6.4}$~K, \opencite{Young2007}), thus a double Gaussian fit was performed to estimate the effect of the blending on the dopplergrams. As the loops selected in \fexii 195~\AA\ were outside the active region core, the \nixvi\ 185.23~\AA\ line intensity is negligible, and does not influence the measured Doppler shift.

To calibrate the zero level of the measured Doppler shifts, we averaged the velocities in small quiet areas near the active region observed during the second raster. We  compared our dopplergrams with the ones published in \inlinecite{DelZanna2008} and found that they were in good agreement. 

For every spectral line, pointing corrections relative to \heii\ 256~\AA\ were taken into account to co-align the EIS data. Furthermore, in order for the different data sets to be comparable, co-alignment was performed among the EIS, TRACE, SOT and MDI data, within the uncertainty of the MDI spatial resolution.

From the aforementioned EIS data we manually selected, at each wavelength, the loops that could be distinguished from the background, along their entire length (Figure~\ref{fig:selected_loops}). For each loop, we recorded  their positions $(x_s,y_s)$ on the intensity image as well as the Doppler shifts $V_{S_z}$, at the same positions, from the Dopplergrams. We applied a boxcar average with a width of 5 to 7 at these line of sight velocities to reduce noise.

The measured positions $(x_s,y_s)$ and line-of-sight velocities $V_{S_z}$ along the loops' length, were then used as input for the computation of the loops' 3D structure following the methodology of Paper~I.
Briefly, the model assumes that a given loop is (i) stationary, (ii) lying in a single plane, and (iii) exhibits plasma flows along its length. The sought-after unknown parameter is the
inclination $\beta$ of the plane with respect to the local solar vertical.
There are two extremes for $\beta$: $\beta_1$, that is the line-of-sight inclination with respect to the local vertical, and a maximum inclination $\beta_2=\beta_1-90\degr$ such that the loop plane is not
submerged below the solar surface.

To find the unknown inclination $\beta$ we first transform the loop's positions and line-of-sight velocities from the sky coordinate system $(x_s, y_s, z_s)$ to the local orthogonal system
$(x_l, y_l, z_l)$, where the $(x_l,y_l)$-plane is tangent to the solar surface and $x_l$ runs parallel to the equator, and then to the loop's system $(x,z)$. Here $\hat{x}$ is defined by the loop's
footpoints, $\hat{z}$ is lying on the loop's plane, and $V(s, \beta)$ is the velocity on the plane, along the loop's length (see Figure~3 of Paper~I for the geometrical setup).

We apply the method by changing the inclination within the interval $[\beta_1, \beta_2]$ in steps of 0.1$\degr$. At each step, we monitor the quantity

$M(\beta) = max{ \Delta V(s,\beta)}$, where $\Delta V(s, \beta) = |V(s_{i+1},\beta) - V(s_i, \beta)|$ is the absolute velocity difference between consecutive pixel positions, i, i+1, along the loop. The optimal $\beta$ is the one
minimizing $M(\beta)$. We further define an uncertainty $\delta \beta$ such that $M(\beta)$ is smaller than, twice the minimum $M(\beta)$-value. In effect this is a $2-\sigma$ uncertainty level;
see Figure~\ref{fig:loop1_betas}c for an example.
Most $\beta$-values within the inspected range are immediately discarded either because large discontinuities appear in $V(s, \beta)$ along the loop or because
$V(s,\beta)$ obtains unrealistically large values. Besides the primary minimum $M(\beta)$ for $\beta \in [\beta_1, \beta_2]$ we sometimes obtain a secondary minimum for $\beta \simeq
\beta_1$. This is also typically discarded as the loop-length solution in this case becomes unrealistically large, larger than the solar radius.

\section{Results} 

\begin{figure}

\centerline{
	      \includegraphics[width=\textwidth]{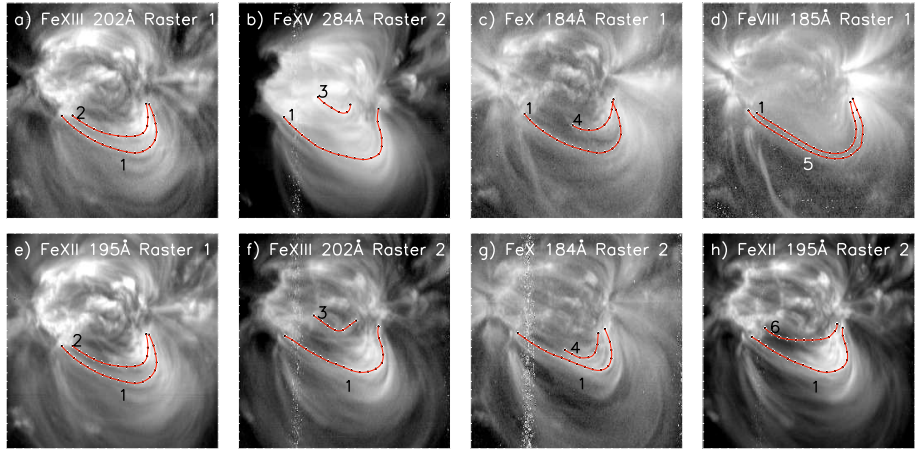}
	    }
\caption{Intensity images, in various spectral lines, from both rasters, on which we indicate the shape of the selected loops. Loop~1 appears in all lines and in all rasters. Loop~2 can be seen in panels a and e, Loop~3 in panels  b and f, Loop~4 in panels c and g, loops~5 and~6 in panels d and h.
	      }
  \label{fig:selected_loops}
\end{figure}

In the studied active region, during both rasters and in all recorded spectral lines, we were able to identify six loops having a sufficient contrast with respect to the background along their entire length. All of them were located in the south part of the active region (solar-Y coordinates less than -200\arcsec). For an additional four loops the method did not work successfully, failing to converge to a continuous velocity solution.

Figure~\ref{fig:selected_loops} shows all studied loops as they appear in intensity images of different spectral lines during the two rasters. Loop~1 was the most prominent, as it appears in both rasters and in all wavelengths  shown in Figure~\ref{fig:selected_loops}. Loop~2 was observed during raster~1 in \fexiii\ 202~\AA\ and \fexii\ 195~\AA\ (Figure~\ref{fig:selected_loops} panels a,e). Loop~3 was observed during raster~2 in \fexv\ 284~\AA\ and \fexiii\ 202~\AA\ (Figure~\ref{fig:selected_loops} panels b,f). Loop~4 was observed in \fex\ 184~\AA\ during both rasters (Figure~\ref{fig:selected_loops} panels c,g). Finally, loops~5 and~6 were observed in \feviii\ 185~\AA\ during raster~1 and in \fexii\ 195~\AA\ during raster~2 respectively (Figure~\ref{fig:selected_loops} panels d,h).

\subsection{Results for Loop 1} 
\subsubsection{Loop reconstruction}

\begin{figure}
\centerline{
    \includegraphics[width=\textwidth]{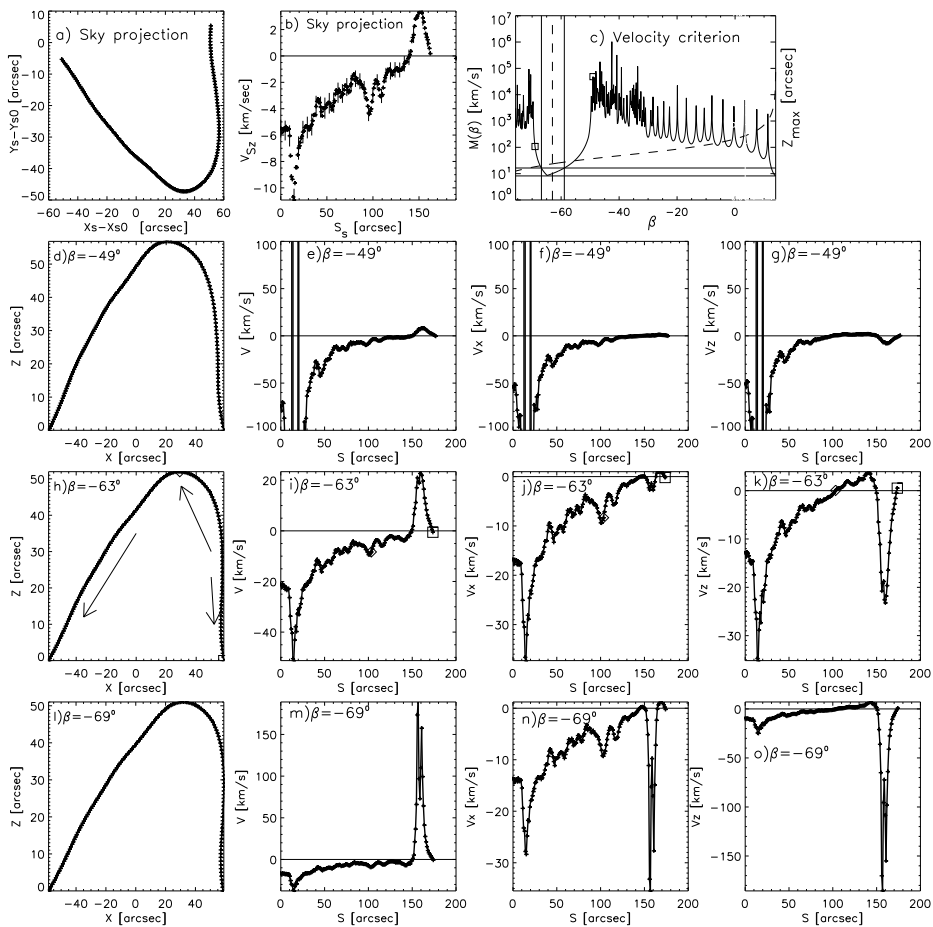}
	    }
\caption{Results for Loop~1 at \fexii\ 195~\AA. First row, panels a and b: Loop position in the sky plane and line-of-sight velocity as a function of the loop's length. panel c, shows $M(\beta)$ (solid line) and $z_{max}(\beta)$, 
maximum loop altitude, (dashed line) as a function of $\beta$. The two vertical lines show the range of optimum solutions while the vertical dashed line indicates the selected $\beta$.  Each 
row of four panels shows calculations for different loop inclinations, $\beta$. These include:
The reconstructed loop in its own plane $(x,z)$ (panels d, h,l), the flow velocity $V$ along the loop  (panels e,i,m), the horizontal component $V_x$ (panels f, j, n) and vertical component $V_z$  (panels g, k, o) as a function of the loop length. The third row is for the optimal value of $\beta =-63\degr$. The diamond indicates the top of the loop and the box indicates the west footpoint. The arrows in panel h) show the direction of the plasma flow. Note the discontinuities of the flow velocity in the second and the fourth rows ($\beta=-49\degr$ and $\beta=-69\degr$). Positive line of sight velocities correspond to redshifts.
	      }
  \label{fig:loop1_betas}
\end{figure}
Figure~\ref{fig:loop1_betas} shows the application of the reconstruction method for Loop~1, from the \fexii\ 195~\AA\ image during raster~2 (see Figure~\ref{fig:selected_loops}, panel h). In the first row, the two first panels show the plane of the sky projection of the loop (panel a) and the Doppler velocity along its length (panel b), where positive values correspond to redshifts. The error bars in panel b are derived from the statistical uncertainties of the Gaussian fit. The data
presented in panels (a) and (b) are the inputs of the reconstruction method. Then, for a wide range of inclinations $\beta$ we computed the loop shape and the velocity along the loop, in the loop reference system $(x,z)$.  Panel c) shows the maximum velocity difference (solid line) $M(\beta)$ and the maximum loop altitude $z_{max}$ (dashed line) as a function of $\beta$. Panel (c) presents a \lq forest\rq\ of large $M(\beta)$ values, of the order of $10^3$ to $10^6$~\kms\ interrupted by a region with values of 10 to 200~\kms. For large $\beta$ values, $M(\beta)$ has another small value region, which corresponds to unphysically large values for $z_{max}$ ($\simeq 10\,000$\arcsec\ ).

Each of the next three rows, has four panels, showing, from left to right, the reconstructed loop on its own plane, the velocity $V$ of the flow along the loop, the horizontal $V_{x}$ and the vertical $V_{z}$ velocity components as a function of the loop length. The second and fourth row show the results for $\beta=-49\degr$ and $-69\degr$ respectively. There, the transformation from the sky plane to the loop plane gives a discontinuity in the computed velocity (panels e, f and g) for $\beta=-49\degr$, and a strong peak in velocities that does not seem physical (panels m, n and o for $\beta=-69\degr$), thus these results are rejected as unphysical. For the third row, where $\beta=-63\degr$, the velocity is continuous (panels i, j and k), and therefore the results are considered acceptable. The examined values of $\beta$ range from $\beta_2=-75\degr$ where the loop plane becomes tangent to the solar surface, to  $\beta_1=14\degr$, where the line-of-sight is inside the loop plane. Negative values of $\beta$ means that the loop is inclined towards the south part of the active region. Let us come back to panel c. At $\beta=-49\degr$ a square  shows the $M(\beta)$ value of $4.8\times 10^4$~\kms\  which corresponds to the subtraction between the most negative values of $v$ seen in panels (e), (f) and (g) at $s=19$~arcsec, where $V$, $V_x$ and $V_z$ resemble with delta functions. Panel c), at $\beta=-69\degr$ a square shows the value of $\simeq 105$~\kms\ which corresponds to the maximum difference values at $s \simeq 155$~arcsec in panels (m), (n) and (o).

The two horizontal lines show the minimum and two times the minimum of $M(\beta)$. These yield a range for $\beta$ (shown between the two continuous vertical lines), defining an uncertainty of $\pm 4\degr$ for the loop's inclination $\beta$. The mean value of $\beta$ in this interval is pointed by the vertical dashed line.

\subsubsection{Interpretation of flows in Loop 1}
The flow velocity $V$, in Figure~\ref{fig:loop1_betas}h, for $\beta=-63\degr$, is negative, with the exception of the west footpoint of the plot. This means that the  velocity flow is towards the east footpoint starting from the upper part of the west leg, through the loop apex as indicated by the two arrows in panel h. There is however also a portion of the loop, near the west footpoint where we observe a downflow indicated by the right arrow in panel h.
In panel $j$, $V_x$, is negative, indicating also a motion towards the east footpoint. The west leg is almost vertical which is why $V_x \simeq 0$. In panel k, $V_z$ also shows a flow towards the east footpoint with negative/down-flows, east of the loop top (diamond), and positive/up-flows west of the loop top. Near the west footpoint, $V_z$ changes sign again, at the same position where also the Doppler shift changes sign (panel b). This would mean that the plasma flows towards the west footpoint in this small section. However, the small Doppler shift values increase the relative error and makes interpretation ambiguous, for this small loop section.

\begin{figure}
\centerline{
	    \includegraphics[width=\textwidth]{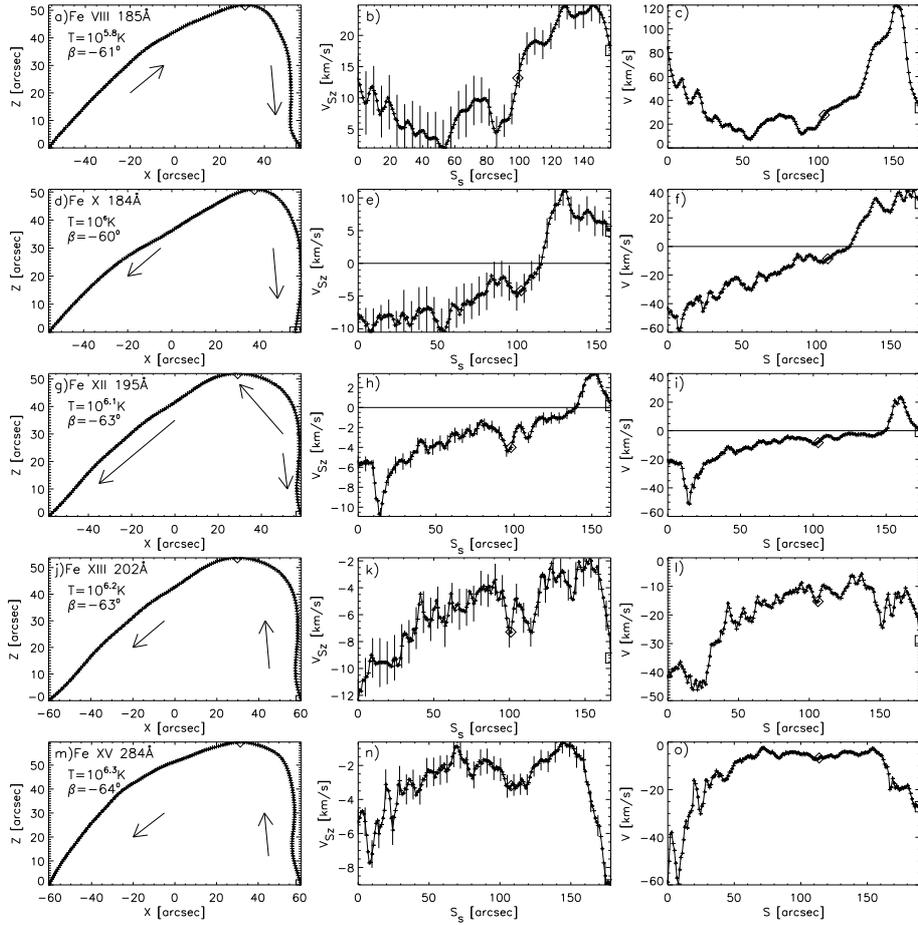}
	    }
\caption{Results for Loop~1 for all spectral lines during Raster~2. First column: The reconstructed loop in its own plane $(x,z)$. Second column: Line-of-sight velocity as a function of the loop's length. Third column: Velocity of the flow as a function of the reconstructed loop's length.
	      }
  \label{fig:loop1_raster2}
\end{figure}

Figure~\ref{fig:loop1_raster2} shows the results for Loop~1, for all spectral lines during raster~2. Each row in Figure~\ref{fig:loop1_raster2} corresponds to a different spectral line and includes three panels. The left panel shows the loop shape on its own plane, the middle panel shows the measured Doppler shift along the loop and the third panel shows the magnitude of the  computed velocity. For the middle panels, positive values mean redshifts and error bars are due to Gaussian statistical uncertainties. 

From the first column of panels we can see that Loop~1 implies similar inclinations in all spectral lines, in the range of $-63\degr$ to $-60\degr$. For the \feviii\ spectral line, we measure redshifts of 5 to 20~\kms\ along the loop (panel b) which for the given loop inclination correspond to unidirectional flow from the east toward the west footpoint (panel c). 
The results for \fex\ spectral line, show that both the Doppler shift (panel e) and the computed velocity (panel f) change sign near the loop top; this indicates a draining motion from the loop top (indicated with a diamond symbol) towards the footpoints. This draining shows velocities of $\simeq 40$~\kms\ at both footpoints. The last two spectral lines, \fexiii\ 202~\AA\ and \fexv\ 284~\AA\ show a similar dynamic behavior.
The measured Doppler shifts are towards the blue (panels k, n) and the computed velocity (panels l, o) shows unidirectional flows from west to east footpoints. This similarity indicates that we observe the same plasma structure in these two spectral lines. The flow of \fexii\ 195~\AA\ in panel (i), has also the characteristics of a unidirectional flow as in panels (l) and (o) but the positive flow at the west footpoint makes this interpretation problematic.
It is worth noting that the flow in the hot lines is in the opposite direction to the flow in the cooler \feviii 185~\AA. Therefore, this may imply that Loop~1 is composed of different strands or background structures.
\hyphenation{in-clu-des}
\begin{table}
\caption{A summary of the 3D reconstruction results for all selected loops. The Table includes: loop number (column 1), spectral line (column 2), raster number (column 3), $\beta$ (column 4), the type of flow, which can be draining, or flow from East to West (E. to W.) or West to East  (column 5), apparent density scale length (column 6) maximum altitude (column 7) and length (column 8). The last three columns, show the velocity mean values along 10\% of the loops length starting respectively from east and the west footpoint and the mean velocity along 10\% of the loop top. The corresponding standard deviation is shown in parenthesis.}
\begin{center}
\begin{tabular*}{\textwidth}{@{}@{\extracolsep{\fill}}llllrrrrr}
\hline\hline
No. Spectral          &Ra-\hspace{.4cm}$\beta \pm \delta \beta$   &flow  & $\lambda_n^{obs}$ & Z & L& \mc{3}{c}{Velocities ($km/s$)} \\ 
\hspace{.4cm} line    &ster\hspace{.7cm}($\degr$)      &     & (\arcsec)& (\arcsec)  & (\arcsec)  & East  & West & Top \\ \hline
1  \feviii\ 185~\AA\  & 1 \hspace{.4cm}$-64 \pm 4$ & E. to W.& 40  &26  & 185   &  74(27)  & 47(14) &47(11) \\
1  \fex\ 184~\AA\     & 1 \hspace{.4cm}$-65 \pm 1$ & Draining& 148 &25  & 182   & -92(19) & 69(22)  &-1(3)\\
1  \fexii\ 195~\AA\   & 1 \hspace{.4cm}$-65 \pm 1$ & Draining& 197 &23  & 176   & -72(15) & 33(12)  &-2(2)\\
1  \fexiii\ 202~\AA\  & 1 \hspace{.4cm}$-65 \pm 1$ & W. to E.& 221 &23  & 176   & -94(18) & -36(14) &-3(2)\\
1  \feviii\ 185~\AA\  & 2 \hspace{.4cm}$-61 \pm 4$ & E. to W.& 126 &26  & 168   &  52(16)  & 76(34)  & 26(7)\\
1  \fex\ 184~\AA\     & 2 \hspace{.4cm}$-60 \pm 4$ & Draining& 230 &25  & 170   & -46(6)  & 36(3)    &-8(3)\\
1  \fexii\ 195~\AA\   & 2 \hspace{.4cm}$-63 \pm 4$ & W. to E.& 253 &24  & 174   & -30(6)  & 9(6)     &-6(1)\\
1  \fexiii\ 202~\AA\  & 2 \hspace{.4cm}$-63 \pm 4$ & W. to E.& 252 &25  & 180   & -40(2) & -17(4)    &-12(1)\\
1  \fexv\ 284~\AA\    & 2 \hspace{.4cm}$-64 \pm 2$ & W. to E.& 272 &26  & 190   & -43(10) & -21(3)   &-6(.5)\\ \hline
2  \fexii\ 195~\AA\   & 1 \hspace{.4cm}$-64 \pm 4$ & Draining& 47  &15  & 129   & -34(4)  & 19(6)    &-4(2)\\
2  \fexiii\ 202~\AA\  & 1 \hspace{.4cm}$-64 \pm 4$ & W. to E.& 102 &15  & 129   & -42(7)  & -9(12)   &-7(2)\\\hline
3  \fexiii\ 202~\AA\  & 2 \hspace{.4cm}$-75 \pm 15$ & W. to E.& 205&4   &  64   & -16(1)  & -11(3)  &-16(1)\\
3  \fexv\   284~\AA\  & 2 \hspace{.4cm}$-61 \pm 6$ & W. to E.& 107 &6   &  55   & -11(1)  & -17(5)   &-7 (1)\\\hline
4  \fex\ 184~\AA\     & 1 \hspace{.4cm}$-69 \pm 7$ & Draining& 66  &7   &  73   & -33(6)  & 31(4)    &-9(5)\\
4  \fex\ 184~\AA\     & 2 \hspace{.4cm}$-65 \pm 7$ & W. to E.& 177 &9   & 65    & -39(4)  & -21(2)   &-6(2)\\\hline
5  \feviii\ 185~\AA\  & 1 \hspace{.4cm}$-67 \pm 0.3$ & E. to W. & 67 &27  & 185   &  30(25)  & 147(32) &31(5)\\
6  \fexii\ 195~\AA\   & 2 \hspace{.4cm}$-68 \pm 5$ & W. to E. &56  &24  & 106   & -14(2) & 3(4)	&-12(2)\\ \hline\hline
\end{tabular*}
\end{center}
\label{tab:loop_full_info}
\end{table}

\subsection{Results for all loops} 
In Table~\ref{tab:loop_full_info} we present a summary of the general characteristics of the flow for all studied loops in all the recorded spectral lines for both rasters. 
As seen in column 4, all the studied loops, in the different spectral lines have a high inclination in the range -60$\degr$ to -75$\degr$ with an average inclination of -64.5$\degr$ and an uncertainty of 3.5$\degr$. 
The inclination measured for Loop~1 has an average of -63.3$\degr$ with an uncertainty  of 1.8$\degr$. Loop 1 inclinations in the first raster have a small scatter, in agreement with the smaller error bars. 
The larger error on $\beta$ is measured for loop 3 (line 12)  and is of $\pm 15\degr$.
In column 5 more than half of the loop (53\%) show unidirectional flow from West to East, 30\% show draining flow from the loop top to the footpoints while the rest of them, all in \feviii\ 185~\AA\ line, have flows from East to West. 
Three of the four cases observed in the \fex\ 184~\AA\ line presented draining motions. 

We note that the dynamics of Loop~1 does not change qualitatively from the first to the second raster, except for the case of the \fexii\ 195~\AA\ line. Loops 4, 5 and 6 also change dynamics from the first to the second raster. 
For the cases with qualitatively similar loop dynamics in both rasters we can argue for a quasi-steady flow along them between 16:05~UT, the time when the first raster started scanning the loop, and 21:31~UT, the time the second raster finished, which is five and a half hours. Moreover, the ratios of the loop lengths (Table~\ref{tab:loop_full_info}, column 8) over the corresponding velocities at the footpoints (columns 9 and 10), give timescales, 90\% of which, are within 25 minutes to 2 hours. These values are close to the duration of the loops rastering (about 1~hour). For the other loops we cannot argue whether or not their dynamical behavior changes between the two rasters. However, as long as the loop magnetic structure is not modified during the raster, 
the applied method should give valuable results for the inclination $\beta$.

The last three columns of Table~\ref{tab:loop_full_info} show the velocity average values along 10\% of the loop's length starting, respectively, from the east footpoint the west footpoint and at the top of the loop, with their standard deviation in parenthesis. In most cases, the east footpoints have larger velocities by, factors from 1.2 to 3. We suggest that the loops appearing in \fexiii\ 202~\AA\ and \fexv\ 284~\AA\ should correspond to the same loop because the calculated dynamics is very similar. Columns 7 and 8 show the maximum altitude from the solar surface and the true loop lengths respectively. Loop~1 is apparently composed of separate strands that appear in different spectral lines as indicated by the different kind of flows found. For each spectral line, the reconstructing method gave values of the loop height (column 7) which vary from 23\arcsec\ to 26\arcsec\ while the loop reconstructed length varies from 170\arcsec\ to 190\arcsec. These variations indicate the uncertainties of the method but could also show the difference between various strands composing the loop.

\subsubsection{Calculation of length scales and Mach numbers}

The hydrostatic pressure scale height, defined in centimeters as ($\lambda_p = 4.7\times 10^9 (T/1MK)$), \cite{Aschwanden2005} for the spectral lines \feviii, 185~\AA, ($10^{5.8}$~K), \fex, 184~\AA, ($10^6$~K), \fexii, 195~\AA, ($10^{6.1}$~K), \fexiii, 202~\AA, ($10^{6.2}$~K), and \fexv, 284~\AA\  ($10^{6.3}$~K), is  41.5\arcsec, 65.7\arcsec,  82.7\arcsec,  104.2\arcsec, and 131.2\arcsec\ respectively, and was computed using the formation temperature of each spectral line given in the parentheses. This scale height is still meaningful for subsonic flows which is the case of our loops as it is shown in the next paragraph. Note that the maximum altitudes (column 7) of all measured loops are smaller than the corresponding pressure scale height. Thus this explains that the tops of loops are dense enough to be bright and detectable.

Column 6  shows the apparent density scale, $\lambda_n^{obs}$ (see \opencite{Aschwanden2005}  p. 84). To calculate $\lambda_n^{obs}$ along the loops, we first corrected the loop intensity from the background contamination. To estimate the background, we represented each loop in images with curvilinear grids, (\opencite{Aschwanden1999} their Figure~6). The x-axis of these images represents the length along the loop so that the loop looks stretched and horizontal. 
For each point $(x_{loop},y_{loop})$ along the loop in the curvilinear grid, we selected pixels lying along the line perpendicular to the loop $(x_{loop},y)$ with intensities $I_{back}(x_{loop},y_{back})<f I_{loop}(x_{loop},y_{loop})$ where factor $f$ is in the range from 0.5 to 0.8. We subtract the average of the background pixels intensities from the corresponding loop pixel intensities to get the corrected intensity along the loop. We then calculate the electron density $n_e(s)$ along the loop as follows. For each loop we use the contribution functions $G(T)$ of the corresponding spectral line from the CHIANTI software \cite{chianti1}. The contribution function for all spectral lines was computed assuming a \opencite{Mazzotta98} ionization fraction and a hybrid abundance for iron \cite{Fludra99}. We assume that the loop width along the line of sight is constant and of $w=2$\arcsec. The electron density is calculated 
as $n_e(s)=\sqrt{\frac{I(s)}{w G(T_{max})} }$. $T_{max}$ is the spectral lines formation temperature, maximizing the contribution function and $s$ is the length along the loop. We performed a fit with an exponential function to the electron density function $n_e(s)$ computed along the loop. The density scale length $\lambda_n^{obs}$  is the derived length from the exponential fit. The resulting relative difference between $\lambda_n^{obs}$ and $\lambda_p/cos(\beta)$ is in the range of 0.01 to 0.8 with an average of 0.46.

The sound velocity, expressed as $v_s\,=\, 0.151\times \sqrt{T}$~\kms, was computed for the different spectral lines formation temperatures to be in the range of 120~\kms\ to 213~\kms. Therefore, the corresponding Mach numbers, are smaller than one with the exception of the west footpoint of loop 5 (row 15 in Table~\ref{tab:loop_full_info}) which is supersonic. For the \feviii, 185~\AA, \fex, 184~\AA, \fexii, 195~\AA, \fexiii, 202~\AA\ and the \fexv, 284~\AA\ lines, the mean Mach numbers computed from the corresponding velocities in columns 9 and 10 (Table~\ref{tab:loop_full_info}) 
are 0.6, 0.3, 0.2, 0.2 and 0.1 respectively and shows that, the larger the line formation temperature the smaller the Mach number.

\subsubsection{Influence of background correction on Doppler shifts}
In order to check the influence of the background on our results, we also performed the loop reconstruction, using loop Doppler shifts calculated from spectral profiles from which we subtracted a background spectral profile \cite{Gontikakis2005}. We got better results when the background points were selected along the same slit position as the corresponding loop points.  
We obtained similar inclinations within $\simeq 3\degr$ compared to the uncorrected case. Qualitatively the flow (draining or unidirectional flow) is the same for nine cases (rows 1,~2,~3,~6,~7,~10,~14,~15,~16 of Table~\ref{tab:loop_full_info}). The calculated flow, in most of the other cases diverged at the west leg of the loops. The west leg of the loops is almost parallel to the slit so that our method to select appropriate background points fails.

\section{Comparing reconstructed with extrapolated coronal-loop shapes}
As an independent test of our 3D reconstructions we used linear force-free magnetic field extrapolations as described by \inlinecite{Alissandrakis1981} to determine whether extrapolated field lines (viewed as loops) that model the reconstructed loops can be found. That active-region magnetic fields are far from a linear force-free configuration (and, in fact, far from a force-free field configuration in general, at least at photospheric heights) is well known (see, e.g., \opencite{Metcalf1995}; \opencite{Georgoulis2004}). However, we are seeking single $\alpha$-values to fit {\it single} loops, with different values corresponding to different loops. In this approach we have better control over the fitting process, that cannot be achieved by a fully-fledged model-dependent nonlinear force-free field extrapolation (see, e.g., \opencite{Schrijver2006}; \opencite{Metcalf2008}). 
To compare the observed and the extrapolated loops we employ the distance modulus $C_l$, introduced by \cite{Wiegelmann2002}, namely, 
\begin{equation}
C_l = {{1} \over {L_{loop}^2}} \int |\mathbf{R}_{loop} - \mathbf{R}_{line}| dl\;\;, 
\label{eq1}
\end{equation}
where $\mathbf{R}_{loop}$, $\mathbf{R}_{line}$ are respective vector positions along the reconstructed loop and extrapolated field line, respectively, and $L_{loop}$ is the loop's length. Integration occurs along the reconstructed  loop's length with respective points considered as points having the same location (in terms of normalized distance from footpoints) along the loop. 
The number of points taken vary with the loops' length: for the smallest loop (loop 3) we used 51 points while for the largest (loops 1 and 5) we used up to 215 points. We only compared loops and extrapolated field lines with footpoints lying within a square with a $30$\arcsec-linear size.

As lower boundaries for the extrapolations we use a Level 1.8 magnetogram from the full-disk  \textit{Michelson Doppler Imager} (MDI; \opencite{Scherrer1995}) onboard SOHO. The course of action is to, first, crop the MDI magnetogram to select the desired photospheric patch and then use the cosine-corrected line-of-sight magnetic field component on the image (sky) plane. We choose this action because de-projecting for the local, heliographic plane would introduce unforeseen uncertainties when similarly de-projecting the EIS images. Point taken, we understand that each approach has weaknesses -- this is an issue that we plan to investigate thoroughly in future studies.  

The linear size of the selected patch is $L=183$ Mm. This determines the extremes of the $|\alpha|$-range to be investigated (e.g., \opencite{Alissandrakis1981}; \opencite{Green2002}), by the formula $|\alpha_{max}|=(2 \pi/L)$. Hence, we scan an $\alpha$-range between $-0.0343$ $Mm^{-1}$ and $0.0343$ $Mm^{-1}$, with a step-size equal to $0.006$ $Mm^{-1}$. 

\begin{figure}[!th]
\centerline{
	      \includegraphics[width=0.9\textwidth]{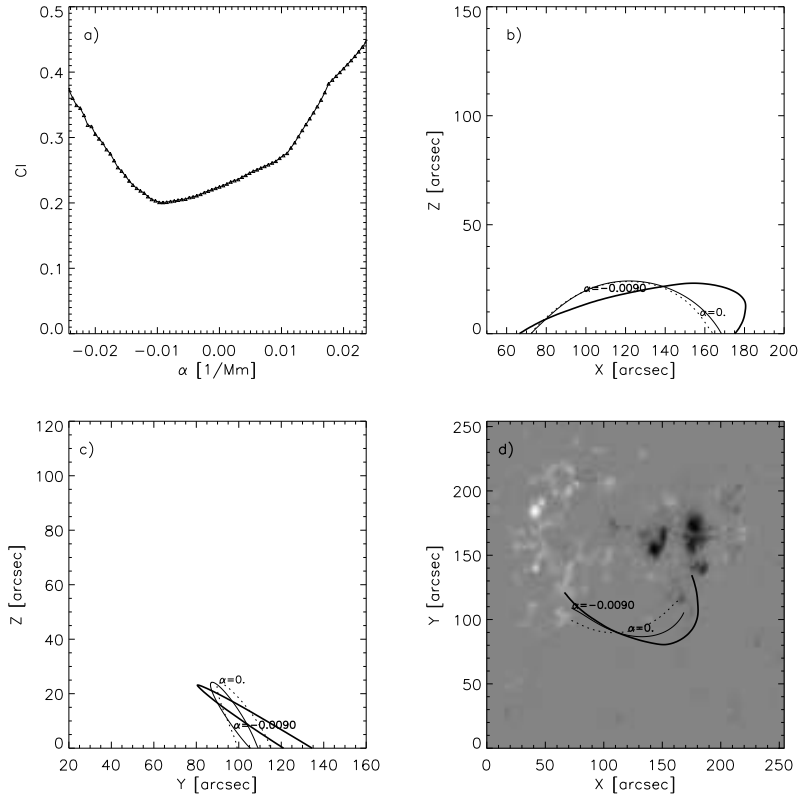}
	    }
\caption{Fitting Loop 1 at \fexii\ 195~\AA\ during raster 1: (a) the $C_l$ modulus along the loop as a function of $\alpha$. A minimum $C_l$ is achieved for $\alpha \simeq -0.009$ $Mm^{-1}$. (b, c) Two different views of the inferred loop (thick solid curve), accompanied by a potential-field extrapolated line ($\alpha =0$; dotted curve) and the "best-fit'' extrapolated field line ($\alpha =-0.009$ $Mm^{-1}$; thin solid curve). (d) The same system, seen from above and projected on the respective SOHO/MDI magnetogram.}
  \label{fig:Cl}
\end{figure}

\begin{table}
\begin{center}
\begin{tabular*}{\textwidth}{@{}@{\extracolsep{\fill}}llllrlcrc}\hline
loop & spectral            &rast.  & $\beta$   &  Z	& $\beta$	 &  Z	& $C_l$	&$\alpha \pm \delta \alpha$            \\ 
num.& line                 & num.  & ($\degr$) & (\arcsec)&  mag.($\degr$)& mag.(\arcsec)& & ($Mm^{-1}$)   \\ \hline
1     & \feviii\ 185~\AA\  & 1     & $-64$     & 26   	& $-55$  	 & 26	&0.23 	&-0.013$\pm$0.006 \\
1     & \fex\    184~\AA\  & 1     & $-65$     & 25   	& $-52$  	 & 25	&0.23 	&-0.012$\pm$0.007 \\   
1     & \fexii\  195~\AA\  & 1     & $-65$     & 23   	& $-51$ 	 & 24	&0.20	&-0.009$\pm$0.006 \\
1     & \fexiii\ 202~\AA\  & 1     & $-65$     & 23   	& $-51$ 	 & 24	&0.24	&-0.009$\pm$0.007 \\
1     & \feviii\ 185~\AA\  & 2     & $-61$     & 26  	& $-50$  	 & 25	&0.13	&-0.006$\pm$0.005 \\
1     & \fex\    184~\AA\  & 2     & $-60$     & 25  	& $-50$  	 & 22	&0.15	&-0.008$\pm$0.004 \\
1     & \fexii\  195~\AA\  & 2     & $-63$     & 24  	& $-50$  	 & 25	&0.15	&-0.006$\pm$0.004 \\
1     & \fexiii\ 202~\AA\  & 2     & $-63$     & 25  	& $-50$  	 & 25	&0.15	&-0.006$\pm$0.004 \\
1     & \fexv\   284~\AA\  & 2     & $-64$     & 26  	& $-48$  	 & 30	&0.14	&-0.003$\pm$0.004 \\
2     & \fexii\  195~\AA\  & 1     & $-64$     & 15  	& $-44$  	 & 13	&0.23	&-0.010$\pm$0.009 \\
2     & \fexiii\ 202~\AA\  & 1     & $-64$     & 15  	& $-47$  	 & 15	&0.23	&-0.015$\pm$0.009 \\
3     & \fexiii\ 202~\AA\  & 2     & $-75$     & 4   	& $-49$  	 & 11	&0.19	&-0.010$\pm$0.004 \\
3     & \fexv\   284~\AA\  & 2     & $-61$     & 6   	& $-50$ 	 & 8	&0.22	&-0.004$\pm$0.006 \\
5     & \feviii\ 185~\AA\  & 1     & $-67$     & 27   	& $-51$ 	 & 27	&0.24   &-0.007$\pm$0.007\\
6     & \fexii\ 195~\AA\   & 2     & $-68$     & 7   	& $-54$ 	 & 12	&0.12	&-0.007$\pm$0.002\\

\hline
\end{tabular*}
\end{center}
\caption{Summary of the results obtained from the magnetic field extrapolation in comparison to the results of the 3D reconstruction. Columns 4 and 5 are the reconstruction results as in Table~\ref{tab:loop_full_info}. Columns 6 to 8 are the extrapolation results for the best fitting magnetic line and column 9 is the calculated best $\alpha$ with its error bar.}
\label{tab:reconstruction}
\end{table}

An example loop fitting, corresponding to Loop 1 at \fexii\ 195~\AA\ is shown in Figure~\ref{fig:Cl}. The minimum $C_l$-value achieved in this case (Figure~\ref{fig:Cl}a; also shown in Table 2) is 
$C_l \simeq 0.2$. Figure~\ref{fig:Cl} also includes the best-fit extrapolated field line (thin solid curve) and a potential-field line (dotted curve) for reference. Notice that, despite the proximity between the best-fit line and the reconstructed loop, the match is not perfect. In particular, the best-fit line shows smaller inclination than the loop and larger distances between the footpoint locations as compared to the distances between the two (loop and field-line) apices. In addition, the $C_l$-curve is quite shallow (as also reported by \opencite{Wiegelmann2002}) so different $\alpha$-values give rise to very similar extrapolated lines. In situations like this it is challenging to assign an uncertainty to each best-fit value. For the purpose of providing an error margin in the best-fit $\alpha$-values, however, we have selected a $10$\%-margin in $C_l$ enclosing the minimum and calculated the standard deviations of $\alpha$-values within this margin. For the example of Figure~\ref{fig:Cl} we find $\alpha = -0.009 \pm 0.006$~Mm$^{-1}$. Even given the uncertainty, therefore, Loop 1 at \fexii\ 195~\AA\ is better fitted by a negative-$\alpha$ field line. 

From the 17 loop cases listed in Table 1 the fitting method gave a minimum (albeit shallow) $C_l$ in 15 cases. The results are summarized in Table 2. From them, we notice that, first, all best-fit $\alpha$-values, including uncertainties, are negative, indicating left-handed twist in the best-fit extrapolated field lines. Second, the distance $Z$ between the solar surface and the reconstructed loop apices is well-reproduced by the fitting. Excluding Loop~3 at \fexiii\ 202~\AA\ (7\arcsec\  of difference) and Loop~6 at \fexii\ 195~\AA\ (5\arcsec\ of difference), the mean amplitude difference in $Z$ between loops and extrapolated field lines is $(1.2 \pm 1.3)$\arcsec, corresponding to $\sim 5.5$\% of the mean loop length. When it comes to loop inclinations, however, extrapolation gives 
{\it consistently} lower inclinations. The mean inclination difference between all reconstructed loops and extrapolated field lines in Table 2 is $14.5\degr \pm 4.3\degr$, or $(22.5 \pm 5)$\% of the reconstructed loop length. The  reason(s) for this systematic difference need to be investigated further. Generally speaking, however, more than one effect may be contributing to this: besides the limited validity of the force-free approximation in the photospheric magnetic fields used as extrapolation boundaries that, however, is not clear whether should yield a systematic lower inclination, (i) the central assumption of loops lying in a single plane may overestimate the loops' inclination as the most highly projected parts of each loop weigh heavier in the reconstruction, and (ii) the extrapolations using the image-plane magnetic field include some projections that may well give rise to an underestimation of the extrapolated field lines' inclination. 

Summarizing, it is indeed a nontrivial exercise to compare reconstructed loops with extrapolated field lines. Previous works (\opencite{Wiegelmann2002}; \opencite{Demoulin2002}; \opencite{Carcedo2003} and others) also reach similar conclusions. Why is this first-order comparison attempted here, then? A rather straightforward answer is because, despite difficulties, (a) we closely match the loop apices' height from the solar surface, and (b) all best-fit lines correspond to negative $\alpha$-values, in statistical agreement with the best-fit $\alpha$-value obtained for the entire NOAA AR 10926 ($\alpha = -0.049 \pm  0.006$ $Mm^{-1}$), as calculated by the method described in \opencite{Georgoulis2007},  using the SOT magnetogram. This implies that the linear force-free method applied separately to each loop consistently yields a dominant left-hand twist in the AR, in agreement with the cruder linear force-free approximation for the entire AR. Even the systematic lower inclination for extrapolated field lines is an effect that 
is interesting to investigate further, as one may be able to assess the validity of the loop reconstruction method {\it vis \'{a} vis} various effects in the extrapolation. We intend to continue and extend these comparisons to clarify these issues. 

\section{Summary and Conclusions}
We reconstructed the 3D geometry of six loops observed by the \textit{Hinode}/EIS spectrograph in five spectral lines during two rasters of the instrument, a total of 17 cases. All loops correspond to NOAA AR 10926. All reconstructed loops have large inclinations with respect to the vertical, in the range of $-60\degr$ to $-75\degr$ and they are all projected to the South of the bright central part of the active region. Moreover, due to pressure scale height effects, large inclinations lead to denser and brighter plasma near the loop top, hence the entire loop stands clearly above the background level. The flows calculated for all loops were subsonic in all spectral lines that  validates the loops description using their hydrostatic scale. 

The best-studied loop is Loop 1, successfully reconstructed in nine images. We find that this 
loop is {\it not a monolithic } structure, as the flows deduced in different spectral lines vary from unidirectonal flow from East to West in the low-temperature \feviii\ line, to draining motion from the top to the footpoints in the intermediate-temperature \fex\ line,  to unidirectional flow from West to East in the high-temperature lines. This being said, 
the computed inclination is very similar in all images, namely $63\degr \pm 3\degr$, that strengthens our assessment of internal structure in the loop in different temperature ranges. 

An independent comparison between the reconstructed loops and extrapolated field lines by means of a linear force-free extrapolation implemented on a case-by-case basis give results that call for additional investigation. We have been able to closely model the height of the loops' apices and we have noticed best-fit lines with a consistent (left-handed in this case) twist in the AR, but extrapolated field lines are consistently less inclined than reconstructed loops. This discrepancy may be due to drawbacks in the reconstruction method, weaknesses in the extrapolation and/or the extrapolated boundary, or a combination of both. Aiming to validate our technique as reliably as possible, we intend to carry out similar investigations in future works, relying on larger statistical samples of reconstructed coronal loops. 

\begin{acks}
This research was supported by research grant 200/740 of the Academy of Athens. CHIANTI is a collaborative project involving researchers at NRL (USA) RAL (UK), and the Universities of: Cambridge (UK), George Mason (USA), and Florence (Italy). 
We would also like to thank the anonymous referee for the valuable comments which improved our paper, as well as the editor for a careful reading of the manuscript. 
\end{acks}

\bibliographystyle{spr-mp-sola}

\bibliography{loop_reconstruction}

\end{article}

\end{document}